\documentclass[
aps,%
12pt,%
final,%
notitlepage,%
oneside,%
onecolumn,%
nobibnotes,%
nofootinbib,%
superscriptaddress,%
showpacs,%
centertags]%
{revtex4}
\def\ba {\begin {array}}
\def\ea{\end {array}}
\def\be {\begin {equation}}
\def\ee {\end {equation}}
\def\bea {\begin {eqnarray}}
\def\eea {\end {eqnarray}}

\begin{document}
\title
{Superintegrability and higher order constants for classical and quantum systems}

\author{\firstname{E. G.}~\surname{Kalnins}}
\affiliation{Department of Mathematics,\\
University
of Waikato, Hamilton, New Zealand.}
\author{\firstname{W.}~\surname{Miller, Jr.}}
\affiliation{
 School of Mathematics, University of Minnesota,\\
Minneapolis, Minnesota, U.S.A.}
\author{\firstname{G.\ S.}~\surname{Pogosyan}}
\affiliation{ Laboratory of Theoretical Physics, 
Joint Institute of Nuclear Research \\
Dubna,Moscow region 141980,Russia
 }

\begin{abstract} We extend recent  work by Tremblay, Turbiner, and Winternitz 
 which analyzes an infinite family of solvable and integrable quantum systems
in the plane, indexed by the positive parameter $k$. Key components of their analysis were to 
demonstrate that there are closed orbits in the corresponding classical system if k is rational, and for a number of examples there are generating quantum symmetries  that are 
 higher order differential operators than two. Indeed they conjectured that for a general class 
of potentials of this type, quantum constants of higher order should exist. We 
give credence to this conjecture by showing that for an even more general class of 
potentials in classical mechanics, there are higher order  constants of   the motion as polynomials in the momenta. Thus these systems are all superintegrable.
\end{abstract}

\maketitle

\section{ Introduction}\label{int1}
Recently Tremblay, Turbiner and Winternitz \cite{TTW1}  studied a  family
of quantum mechanical systems in two dimensions with 
Hamiltonian 
\be\label{TTWham}
H=\partial ^2_r+\frac{2}{ r}\partial _r+\frac{1}{ r^2}\partial ^2_\theta +ar^2+ 
\frac{b}{ r^2\cos ^2k\theta } +\frac{c}{ r^2\sin ^2k\theta }.\ee
They showed that for  $k$  an  integer  this system is superintegrable, as is the corresponding classical analog.  We will prove that if 
$k$ is rational  then the corresponding classical problem with Hamiltonian 
\be\label{KMP1}
H=p^2_r+\frac{1}{ r^2}p^2_\theta +ar^2+ \frac{b}{ r^2\cos ^2k\theta } 
+\frac{c}{r^2\sin ^2k\theta }\ee
is also superintegrable. Indeed,  it has two functionally independent constants of the motion
that are polynomial in the momenta. One of these is always of second order 
corresponding to separation of variables in polar coordinates, viz 
\be\label{angmom}
L_2=p^2_\theta +\frac{b}{ \cos ^2k\theta } +\frac{c} {\sin ^2k\theta }.\ee
It is the other generating constant of the motion on which we will concentrate.

To make this all explicit we consider the case $k=2$ first and then give a 
general construction for rational $k$. Following this  we consider a second potential
\be\label{newpot}
V=a \frac{(x+iy)^{k-1}}{ (x-iy)^{k+1}}\ee
where we also demonstrate superintegrability for $k$ rational. These results strongly
suggest that   corresponding quantum systems are superintegrable.

\section{The potential of Tremblay,Turbiner and Winternitz}\label{sec2}
We look first  at the $k=2$ case.  In Cartesian coordinates the 
classical Hamiltonian is 
\be\label{classicalk=2} H=p^2_x+p^2_y+a(x^2+y^2)+b \frac{(x^2+y^2)}{ (x^2-y^2)^2} +c 
\frac{(x^2+y^2)}{ x^2y^2}.\ee
There are two fundamental constants of the motion:
$$ C_1=(xp_y-yp_x)^2+4b \frac{x^2y^2}{ (x^2-y^2)^2} + c \frac{(x^4+y^4)}{ x^2y^2},$$
$$C_2=(p^2_x-p^2_y)^2+[2ax^2+ 2b \frac{(x^2+y^2)}{ (x^2-y^2)^2} -2c 
\frac{(x^2-y^2)}{ x^2y^2} ]p^2_x$$ 
$$+[-4axy+ 8b\frac {xy}{ (x^2-y^2)^2} ]p_xp_y+[2ay^2+ 2b 
\frac{(x^2+y^2)}{ (x^2-y^2)^2} +2c \frac{(x^2-y^2)}{ x^2y^2} ]p^2_y$$
$$+a^2(x^2-y^2)^2+ \frac{b^2}{ (x^2-y^2)^2} + c^2 \frac{(x^2-y^2)^2}{ x^4y^4}+8ab 
\frac{x^2y^2}{r (x^2-y^2)} +2\frac {bc}{ x^2y^2}.$$

The structure of the symmetry algebra is not worked out in \cite{TTW1}. We give the structure here. 
We set $R=\{C_1,C_2\}$. The Poisson algebra relations are 
$$\{C_1,R\}=32(H^2-2C_2)C_1-64(b+2c)C_2+64(b-c)H^2-128abC_1-128ab(b+2c)$$ 
$$\{C_2,R\}=32C_2(C_2-H^2)+128aC_1H^2-384a^2C^2_1+128abC_2-64(b+4c)aH^2$$
$$+256a^2(
2c-b)C_1
+128a^2(b^2+40c^2+20bc).$$
There is a Casimir constraint  
$$R^2=64C_1C_2(H^2-C_2)-64bH^4+128(b-c)C_2H^2-64(b+2c)C^2_2-128aC^2_1H^2+$$ 
$$256a^2C^3_1-256abC_1C_2+128a(b+4c)H^2C_1+256a^2(b-c)C^2_1-256ab(b+2c)C_2$$ 
$$+256a(7bc+b^2-2c^2)H^2-256a^2(b^2+4c^2+20bc)C_1-256a^2(2c+b)(b^2+16bc-4c^2).$$

From these relations we see that we have a closed Poisson algebra in the same 
sense as found for  many well known superintegrable systems in two dimensions
that are of second order \cite{KKM2, KKMP, DASK2005}. In addition we can look for interesting one 
variable models of this algebra \cite{KMP}. One  such model is 
obtained by choosing $C_1=c$. If we do this then 
$$C_2=\exp (\sqrt{ -C-b-2c}\beta )+\frac{1}{ 2}E^2-2ab- 
\frac{E^2(4c-b)}{2(C+b+2c)}$$
$$-\frac{1}{ 16} 
\frac{(4Cc-4c^2-C^2+16bc)(4ab+9ac+4aC-E^2)^2}{(C+b+2c)^2}
 \exp (-\sqrt{-C-b-2c}\beta ).$$
Here $\beta $ is the variable conjugate to $c$. 

For  the 
quantum analogue of this system the Hamiltonian is 
$$H=\partial ^2_x+\partial ^2_y+a(x^2+y^2)+b \frac{(x^2+y^2)}{ (x^2-y^2)^2} +c 
\frac{(x^2+y^2)}{ x^2y^2}$$
with quantum symmetries 
$$C_1=(x\partial _y-y\partial _x)^2+4b\frac{x^2y^2}{ (x^2-y^2)^2} + c 
\frac{(x^4+y^4)}{ x^2y^2},$$
$$C_2=(\partial ^2_x-\partial ^2_y)^2+(2ax^2 + 2b \frac{(x^2+y^2)}{ (x^2-y^2)^2} -
2c\frac {(x^2-y^2)}{ x^2y^2})\partial ^2_x +$$
$$(-4axy +\frac {8bxy}{ (x^2-y^2)^2})\partial _x\partial _y + (2ay^2 + 2b 
\frac{(x^2+y^2)}{ (x^2-y^2)^2} + 2c\frac {(x^2-y^2)}{ x^2y^2})\partial ^2_y$$
$$+(2ax - \frac{4c}{ x^3})\partial _x+(2ay - 
\frac{4c}{ y^3})\partial _y+a^2(x^2-y^2)^2 + \frac{b^2}{ (x^2-y^2)^2} + 
\frac{c^2(x^2-y^2)}{ x^4y^4}^2+$$
$$
8ab \frac{x^2y^2}{ (x^2-y^2)^2} + \frac{2bc}{ x^2y^2} + 6c(\frac{1}{ x^4} + 
\frac{1}{ y^4}).$$
For these quantum operators there is corresponding closure given by the 
formulas ($\{\cdot,\cdot\}$, $\{\cdot,\cdot,\cdot\}$ are operator symmetrizers) 
$$R=[C_1,C_2],$$
$$[C_1,R]=32C_1H^2-32\{C_1,C_2\}+64(b-c+2)H^2-64(b+2c+4)C_2-128a(b+1)C_1$$
$$-128a(b^2+2bc+4b+6c+4),$$
$$[C_2,R]=32C^2_2-32H^2C_2+128aC_1H^2+128a(b+1)C_2-64a(b+4c+6)H^2-384a^2C^2_1$$
$$-256a^2(b-2c-14)C_1+128a^2(-8+8c+18b+20bc+b^2+4c^2).$$
There is also the Casimir operator
$$R^2=32H^2\{C_1,C_2\}-\frac{32}{ 3}\{C_1,C_2,C_2\}-(64\beta +128\gamma +\frac{2816}{ 3})C^2_2+(128(\beta -\gamma )+\frac{2816}{ 3})H^2C_2$$
$$-(192+64\beta )H^4-128\alpha H^2C^2_1-128\alpha (\beta +1)\{C_1,C_2\}+\frac{128}{ 3}\alpha (12\gamma +3\beta +50)H^2C_1+256\alpha C^3_1$$
$$-\frac{256}{ 3}\alpha (44\beta +44+18\gamma +3\beta ^2+6\beta \gamma )C_2-256\alpha ^2(2\gamma -\beta 46)C^2_1+\frac{256}{ 3}\alpha (42+22\gamma +40\beta
$$
$$+3\beta ^2-6\gamma ^2+21\beta \gamma )H^2-\frac{256}{ 3}\alpha ^2(152\gamma -88+182\beta +3\beta ^2+12\gamma ^2+60\beta \gamma )C_1$$
$$+\frac{256}{ 3}(280\gamma ^2-80\beta ^2+24\gamma ^3+320\gamma +48\beta -4\beta \gamma -84\beta \gamma ^3-54\gamma \beta ^2-3\beta ^2+28).$$

We now prove our central result that the classical Hamiltonian has, in 
addition to the obvious  second order constant of the motion, another independent
 constant of the motion that is polynomial in the momenta. We 
use the results of a previous paper \cite{KKMPog02}. A similar approach was used by Verrier and Evans \cite{Evans2008a}.
 For the general potential we have in 
polar coordinates 
$$ V=\alpha r^2+ \frac{\beta }{ r^2\cos ^2(k\theta )}+ 
\frac{\gamma }{ r^2\cos ^2(k\theta )}.$$
In terms of the new variable $r=e^R$ the Hamiltonian assumes the form 
$$H=e^{-2R}(p^2_R+p^2_\theta +\alpha e^{4R}+\frac{\beta }{ \cos ^2(k\theta )}+ 
\frac{\gamma }{ \sin ^2(k\theta )}).$$
Applying the method of \cite{KKMPog02}
to find the extra invariants we  first need to construct a 
function $M(R,p_R)$ which satisfies $\{M,H\}=e^{-2R}$, or
$$(-4\alpha e^{4R}+2He^{2R})\partial _{p_R}M+2p_R\partial_R M=1.$$
This equation has a solution  
$$M=\frac{i}{ 4\sqrt{ L_2}} B$$
where 
$$
\sinh{ B} =i \frac{(2L_2e^{-2R}-H)}{ \sqrt{ H^2-4\alpha L_2}},\quad
\cosh{B }= \frac{2\sqrt{ L_2}e^{-2R}p_R}{ \sqrt{ H^2-4\alpha L_2}},$$
and 
$$L_2=p^2_\theta +\frac{\beta }{ \cos ^2(k\theta )}+ 
\frac{\gamma }{ \sin ^2(k\theta )}$$
and we also have the relation (which we can use to consider $M$ as a function of $R$ alone):
$$p^2_R+L_2+\alpha e^{4R}-e^{2R}H=0.$$
We now need to find the corresponding function $N(\theta ,p_\theta )$ which
 satisfies $\{N,H\}=e^{-2R}$, or $$(\frac{\beta }{ \cos ^2(k\theta )}+ 
\frac{\gamma }{ \sin ^2(k\theta )})'\partial _{p_\theta }N-2p_\theta \partial _\theta N=1$$
where the prime denotes differentiation with respect to $\theta$. This equation has a
solution 
$N= - \frac{i}{ 4\sqrt{ L_2}k} A$
where  
$$
\sinh{A }= 
i\frac{-\gamma +\beta -L_2\cos (2k\theta )}{ \sqrt{(L_2-\beta -\gamma )^2-4\beta \gamma }},\quad
\cosh{ A }= 
\frac{\sqrt{ L_2}\sin (2k\theta )p_\theta }{\sqrt{ (L_2-\beta -\gamma )^2-4\beta \gamma }}.$$
The constant of the motion is $M-N$, and, since it is constructed such that $\{M-N,L_2\}\ne 0$, it is functionally independent of $L_2$, \cite{KKMPog02}.

From these expressions for $M$ and $N$  we see that if $k$ is rational,
$k=\frac{p}{ q}$ ( where $p, q$ are relatively prime integers) then  
$$\sinh(-4ip\sqrt{ L_2}[N-M])=-\sinh(qA+pB),\quad \cosh(-4ip\sqrt{ L_2}[N-M])=\cosh(qA+pB),$$
each give rise to a classical constant of the motion which is polynomial in the 
momenta.

This can be  seen by  observing that these constants of the motion  can be expressed as factor functions of $L_2$ and 
$H$ times a factor which is polynomial in the canonical momenta, via  the relations
$$ (\cosh x\pm\sinh x)^n=\cosh nx\pm\sinh nx,\  \cosh (x+y)=\cosh x\cosh y+\sinh x\sinh y,$$
$$   \sinh (x+y)=\cosh x\sinh y+\sinh x\cosh y.$$ 
In particular,
$$ \cosh nx=\sum_{j=0}^{[n/2]} \left( \ba{c}n\\ 2j\ea\right) \sinh ^{2j}x\ \cosh^{n-2j}x,$$
$$  \sinh nx=\sinh x\sum_{j=1}^{[(n=1)/2]} \left( \ba{c}n\\ 2j-1\ea\right) \sinh ^{2j-2}x\ \cosh^{n-2j-1}x.$$

Thus if $p,q$ are both odd then 
$$\cosh (qA+pB)=\frac{ C}{[ \sqrt{(L_2-\beta -\gamma )^2-4\beta \gamma }]^q[\sqrt{ H^2-4\alpha L_2}]^p},$$
$$ \sinh (qA+pB)=\frac{\sqrt{L_2}\ D}{[ \sqrt{(L_2-\beta -\gamma )^2-4\beta \gamma }]^q[\sqrt{ H^2-4\alpha L_2}]^p},$$
where $C, D$, are polynomial constants of the motion of orders  $2(p+q), 2(p+q)-1$, respectively. If one of $p, q$ is odd and the other even, then
$$\cosh (qA+pB)=\frac{ \sqrt{L_2}\ C'}{[ \sqrt{(L_2-\beta -\gamma )^2-4\beta \gamma }]^q[\sqrt{ H^2-4\alpha L_2}]^p},$$
$$\sinh (qA+pB)=\frac{ D'}{[ \sqrt{(L_2-\beta -\gamma )^2-4\beta \gamma }]^q[\sqrt{ H^2-4\alpha L_2}]^p},$$ where $C', D'$, are polynomial constants of the motion of orders  $2(p+q)-1, 2(p+q)$, respectively.
 ( We  have in fact produced two extra constants of the motion
whose degree differs by 1. This is easily understood by realizing that we have 
one extra constant and its Poisson bracket with $L_2$.)
For example, if $p=1,\ q=2$ we have (with ${\cal L}=-\gamma+\beta-L_2\cos\theta$),
$$\cosh(2A+B)=\frac{2\sqrt{L_2}\left[ (e^{-2R}p_R(L_2\sin^2\theta p_\theta^2-{\cal L}^2)-\sin\theta(2L_2e^{-2R}-H){\cal L}\right]}{[(L_2-\beta-\gamma)^2-4\beta\gamma]\sqrt{H^2-4\alpha L_2}},$$
$$\sinh (2A+B)=\frac{\left[ (2L_2e^{-2R}-H)(L_2\sin^2\theta p_\theta^2-{\cal L}^2)+4L_2\sin\theta e^{-2R}p_\theta p_R {\cal L}\right]}{[(L_2-\beta-\gamma)^2-4\beta\gamma]\sqrt{H^2-4\alpha L_2}}.
$$
The bracketed quantities in the numerators are 5th and 6th order constants of the motion, respectively.

 \section{ A new potential}\label{sec3}
In addition to the potentials of Turbiner et.\ al.\ there is another family 
based on the same principle. To illustrate the properties of this family 
consider 
\be\label{k=3}H=p^2_x+p^2_y+ a\frac {(x+iy)^6}{ (x^2+y^2)^4}.\ee
This Hamiltonian  admits  three  constants of the motion:
$$K_1=(p_x-ip_y)^3- \frac{a}{ (x-iy)^3}[-(iy+3x)p_x+(-ix+3y)p_y],$$
$$K_2=(xp_y-yp_x)(p_x-ip_y)^3+ \frac{a}{ (x-iy)^3} 
[(3x^2+3ixy-2y^2)p^2_x-(2x^2+3ixy-3y^2)p^2_y-$$
$$i(x+3iy)(iy+3x)p_xp_y- a^2\frac{(x+iy)^3}{ (x-iy)^6},$$
$$K_3=(xp_y-yp_x)^2+ 2ia \frac{y(3x^2-y^2)}{(x-iy)^3}.$$
The Poisson algebra relations are 
$$\{K_1,K_2\}=3iK^2_1,\ \{K_1,K_3\}=6iK_2,$$
$$\{K_2,K_3\}=6iK_1(K_3+a),$$
together with the constraint 
$$K^2_1K_3-K^2_2+a(K^2_1-H^3)=0.$$
There is also a sixth order symmetry $K^2_1$. There are a number of one
variable models to consider for this Poisson algebra which help with the formulation of corresponding 
quantum problems viz.
$$
(1):\ K_3=c,\ K_1=-\sqrt{ \frac{aE^3}{ a+c}}\cos (6\sqrt{ c+a}\beta ),\ 
K_2=-i\sqrt{ aE^3}\sin (6\sqrt {c+a}\beta ).$$
$$(2):\ K_1=c,\ K_2=3ic^2\beta  ,\ K_3=-8c^2\beta ^2 + \frac{aE^3}{ c^2}- a.$$
$$ (3):\ K_2=c ,\ K_1=\frac{i}{ 3\beta } ,\ K_3=-9(c^2+aE^3)\beta ^2-a.$$
We see that (1) indicates a realization of the quantum operators in terms of difference 
operators and (2) and (3) a realization in terms of differential operators.

Proceeding to the quantum analogue we obtain the  operators  
$$K_1=(\partial _x-i\partial _y)^3+ 
\frac{a}{ (x-iy)^3}[-(iy+3x)\partial _x+(3iy+x)\partial _y],$$
$$K_2=(x\partial _y-y\partial _x)(\partial _x-i\partial _y)^3+ 
\frac{a}{ (x-iy)^3}[i(2y^2-3ixy-3x^2)\partial ^2_x-(3iy+x)(iy+3x)\partial _x
\partial _y+$$
$$i(2x^2+3ixy-3y^2)\partial ^2_y-2i(3iy+x)\partial _x-2(iy+3x)\partial _y-8i]+ia
^2 \frac{(x+iy)^3}{ (x-iy)^6},$$
$$K_3=(x\partial _y-y\partial _x)^2+2iay\frac {(-y^2+3x^2)}{ (x-iy)^3},$$
$$H=\partial ^2_x+\partial ^2_y+a\frac {(x+iy)^6}{(x^2+y^2)^4},$$
with the commutation relations 
$$[K_1,K_2]=3iK^2_1,\ 
[K_1,K_3]=6iK_2-9K_1,$$
$$[K_2,K_3]=3i\{K_1,K_2\}+i(27+6a)K_1+9K_2,$$
and the analogue of the constraint 
$$\frac{1}{ 2}\{K_1,K_1,K_3\}-3K^2_2-i\frac{9}{ 2}\{K_1,K_2\}+(\frac{63}{ 2} 
+3a)K^2_1-3aH^3=0.$$

A one dimensional model of this algebra is  

$$K_1=-\frac{i}{ 3x} ,\ K_2= \frac{d}{ dx}.$$
$$K_3=-9x^2 \frac{d^2}{ dx^2} -27x \frac{d}{ dx} -(9+a+9aE^3x^2).$$

We now look at the question of what the constants of the motion might be for 
more general potentials of  type (\ref{k=3}). We consider the potentials  
\be\label{generalk} V=a \frac{(x+iy)^{k-1}}{ (x-iy)^{k+1}}.\ee

As in the previous example it is convenient to pass to variables $R$ and 
$\theta$. In these coordinates the Hamiltonian and the obvious constant of the motion $L$ assume the form 
$$H= 
\frac{(p^2_R+p^2_\theta +a e^{2ik\theta })}{ e^{2R}},\quad L=p^2_\theta +a e^{2ik\theta},\quad V=ae^{2ik\theta-2R}$$

Using the the usual prescription for obtaining the extra constant we need to look 
for solutions of 
$$2He^{2R}\partial _{p_R}M+2p_R\partial _RM=1.$$
and 
$$-2ia e^{2ik\theta }\partial _{p_\theta }N + 
2p_\theta \partial _\theta N=1$$
The  new constant is then 
$M-N$.
If $k$ is an integer it is convenient to consider the solution such that 
$-ik\sqrt {L}(M-N)=A+kB$
where 
$$
\sinh{ A}=\frac{p_\theta }{ \sqrt {a}}e^{-ik\theta },\ 
\cosh{ A}=\sqrt {\frac{L}{ a}}e^{-ik\theta },$$
$$\sinh{ B}=\frac{ip_R}{ \sqrt{ H}} e^{-R},\ 
\cosh{ B}=\sqrt {\frac{L}{ H}}e^{-R}.$$

If $k={p\over q}$ is rational then we consider
$\sinh(qA+pB)$
and  $\cosh(qA+pB)$ in order to obtain extra constants of the classical motion. For example, in the special case 
$k=2$ we obtain  
$$\sinh(A+2B)=\frac{1}{ \sqrt{ a}H}(2(p_\theta +ip_R)L-p_\theta He^{2R})e^{-2R}e^{-2i\theta },$$
$$\cosh(A+2B)=\frac{\sqrt{L}}{ \sqrt{ a}H}(2L-He^{2R}+2ip_\theta p_R)e^{-2R}e^{-2i\theta },$$
where we need only consider these hyperbolic functions multiplied by $\sqrt{ a}H$ in the first case and $\sqrt{ a}H/\sqrt{L}$ in the second,
to obtain the polynomial solutions we seek.

\section{ Conclusion}\label{sec4}
We have shown that for $k$ rational all the classical mechanical systems (\ref{TTWham})
 admit one second order constant of the motion as well as two others of higher 
order  as  polynomials in the momenta. This proves superintegrability and supports recent 
studies by Tremblay,Turbiner and Winternitz \cite{TTW1,TTW2} of the potentials with $k$ 
rational where it has been demonstrated that all the orbits are closed. We also studied 
 a new class of systems (\ref{generalk})  and showed 
that again the systems are superintegrable and demonstrated  how to find  a maximal set of constants polynomial in the 
momenta. We provided  some information about the structure of the symmetry algebras associated with all these systems.

\end{document}